\documentclass[aps,prl,twocolumn]{revtex4-1}
\usepackage{amssymb}
\usepackage{amsmath,bm,mathbbol}
\usepackage{graphicx}
\usepackage{bbm}
\usepackage[bottom]{footmisc}
\usepackage{multirow}

\usepackage{hyperref}
\hypersetup{
  colorlinks   = true, 
  linkcolor    = blue, 
  citecolor   = blue 
}

\begin{document}
\title{Scaling description of frictionless dense suspensions under inhomogeneous flow}
\author{Bhanu Prasad Bhowmik}
\author{Christopher Ness}
\affiliation{School of Engineering, University of Edinburgh, Edinburgh EH9 3JL, United Kingdom}

\begin{abstract}
Predicting the rheology of dense suspensions under 
inhomogeneous
flow is crucial in many industrial and geophysical applications,
yet the conventional `$\mu(J)$' framework is limited to homogeneous conditions
in which the shear rate and solids fraction are spatially invariant.
To address this shortcoming,
we use particle-based simulations of frictionless dense suspensions to derive new constitutive laws that unify the rheological response under both homogeneous and inhomogeneous conditions.
By defining a new dimensionless number associated with particle velocity fluctuations
and combining it with the viscous number, the macroscopic friction and the solids fraction,
we obtain scaling relations that collapse data from homogeneous and inhomogeneous simulations.
The relations allow prediction of the steady state velocity, stress and volume fraction fields using only knowledge of the applied driving force.
\end{abstract}
\maketitle

\paragraph{Introduction.}
Dense suspensions are an important class of soft matter system comprising Brownian or non-Brownian particles mixed roughly equally by volume with viscous fluid~\cite{NessReview}.
Their rheology attracts sustained interest from physicists due to the manifold complex phenomena that arise with apparently simple constituents~\cite{stickel2005fluid,jamali2020rheology}.
These include non-equilibrium absorbing state transitions \cite{DavidPineAbsorbingState},
shear thickening \cite{thickening2},
thinning~\cite{de1985hard},
and yield stress behaviour~\cite{richards2020role}.
As well as being of fundamental interest,
characterising this complexity is key to the extensive use of dense suspensions in various formulation and processing industries. 

A useful model with which to build rheological understanding is the
non-Brownian suspension~\cite{guazzelli2018rheology},
an especially appealing system when one considers the case of inertialess hard spheres.
By analogy to dry granular systems~\cite{jop2006constitutive},
a recent study successfully obtained constitutive laws for this system~\cite{BoyerGuazelliPouliquenPRL2011},
confirming their rate-independence and
finding one-to-one relations between the volume fraction $\phi$
and each of two dimensionless rheological quantities,
the viscous number $J = \eta\dot{\gamma}/P$
and the macroscopic friction coefficient $\mu = \sigma_{xy}/P$.
Here $\eta$ is the suspending liquid viscosity,
$\dot{\gamma}$ is the shear rate,
$P$ is a measure of the particle contribution to the normal stress,
and $\sigma_{xy}$ is the shear stress.
This important result, the so-called $\mu(J)$-rheology, forms the basis of subsequent models that introduce rate-dependence through additional stress scales~\cite{shearThickeningCates,guy2018constraint}.

The applicability of $\mu(J)$ becomes limited when considering inhomogeneous flows
in which $\dot{\gamma}$ varies spatially~\cite{Migration1, Migration2, GillissenNessPRL2020}.
In particular,
the lower limit of $\mu$ (which we denote $\mu_J$)
is non-zero in all homogeneously flowing systems irrespective of the particle-particle friction coefficient $\mu_p$~\cite{da2005rheophysics,chialvo2012bridging,ChealNessJoR2018}
but can by construction vanish when mechanical balance dictates sign changes in $\sigma_{xy}$ such as along pipe centrelines.
In such scenarios regions that would otherwise be jammed (\emph{i.e.} with $\mu<\mu_J$ and $J=0$) can have non-zero $\dot{\gamma}$ thanks to facilitation by nearby flowing regions~\cite{TigheNonLocalPRL, PolluiqenNonLocalEffect}.
This non-local effect has been extensively studied in amorphous solids~\cite{BocquetNonLocalNature} and dry granular systems~\cite{KamrinKovalPRL2012},
often by formulating a fluidity field with diffusive behaviour characterised by an inhomogeneous Helmholtz equation.
Microscopically it is conceptualized that the fluidity originates from an activated process
that diffuses through the system in a cooperative way controlled by an inherent length scale~\cite{BocquetNonLocalNature, BocquetNonLocalPRL, KamrinKovalPRL2012, TigheNonLocalPRL, TrulssonNolocal}.
Recent works in dry granular matter~\cite{ZhangKamrinPRL2017,KimKamrin2020PRL,gaume2020microscopic}
interpret the fluidity 
in terms of particle velocity fluctuations $\delta u$ and density $\rho$,
defining a fourth dimensionless quantity $\Theta=\rho \delta u^2/P$
and seeking constitutive relations linking it to $\phi$, $\mu$
and $I$~\cite{jop2006constitutive} (the dry counterpart to $J$).
This successfully collapses data from homogeneous and inhomogeneous simulations onto a master curve, but is limited in that the $\Theta$ fields required to make predictions thereafter must be obtained by simulation.
Naturally such findings raise the question of whether similar constitutive equations exist to unify homogeneous and inhomogeneous dense suspension rheology.

Here we use particle-based simulation~\cite{CundallAndStrackDEM} to model dense suspensions under homogeneous and inhomogeneous conditions,
achieving the latter through an imposed Kolmogorov flow following the approach of~\citet{TigheNonLocalPRL}. 
We seek to unify the rheology under both sets of conditions by first defining a dimensionless suspension temperature based on particle velocity fluctuations, as $\Theta = \eta\delta u/aP$,
analogous to the granular temperature~\cite{KimKamrin2020PRL},
and then obtaining relations among the four dimensionless numbers $\phi$, $J$, $\mu$ and $\Theta$.
Although the $\mu(J)$ framework was devised based on frictional millimetric grains,
recent experiments demonstrate it is nonetheless applicable to frictionless ones~\cite{etcheverry2023capillary},
and we focus here on the latter.
Doing so we find scalings that can collapse homogeneous and inhomogeneous rheology data onto a set of master curves
that can then be used to predict the rheology of other flow types.

\begin{figure*}
\includegraphics[trim = 0mm 135mm 0mm 0mm, clip,width=0.95\textwidth,page=1]{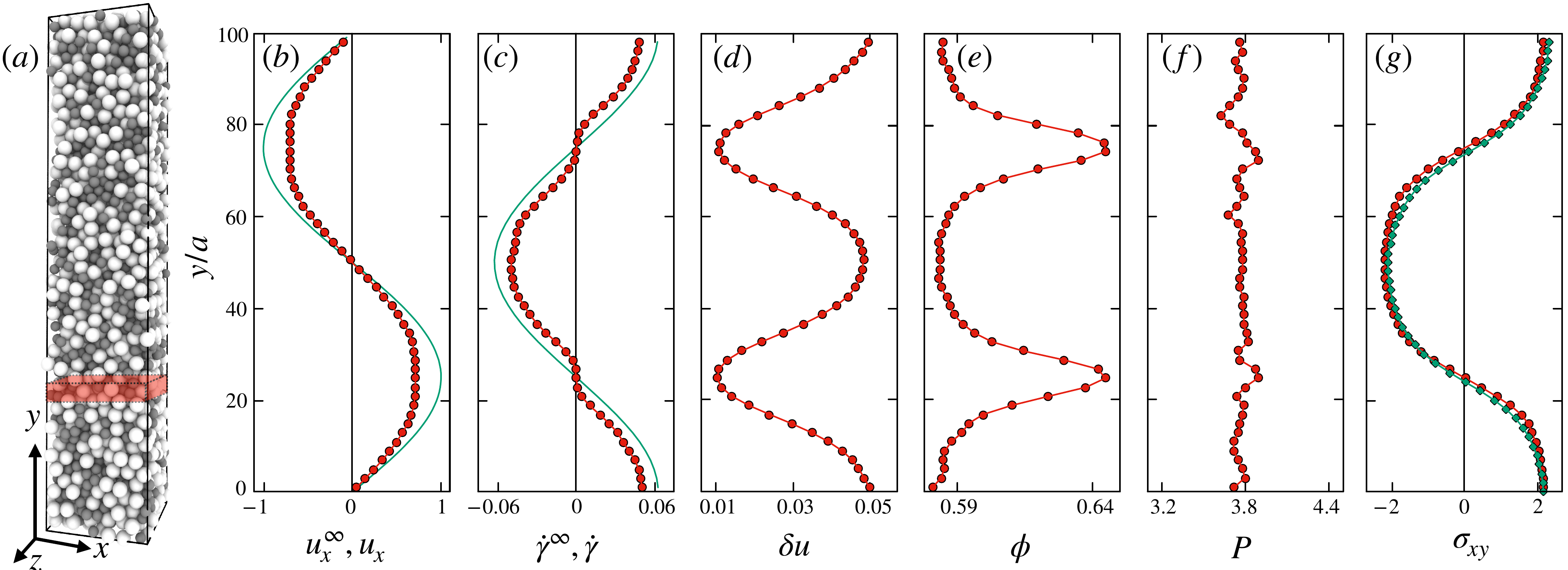}
\hspace*{.7cm}
\caption{
Inhomogeneous flow of a frictionless dense suspension.
Shown are
(a) a typical configuration of the system for $\bar{\phi} = 0.60$, with the red region highlighting a coarse-graining box;
and the steady-state profiles in $y$ of
(b) the $x$-components of the externally applied liquid velocity field $u_x^{\infty}$ (green line) and the coarse-grained velocity field of the particles ${u}_x$ (red points). Velocity is presented here in units of $\kappa$;
(c) the expected shear rate for a Newtonian fluid $\dot{\gamma}^\infty = \partial u_x^{\infty}/\partial y$ (green line) and the measured shear rate $\dot{\gamma}$ (red points),
both in units of $\kappa/a$;
(d) the velocity fluctuations $\delta u$ in units of $\kappa$;
(e) the local volume fraction $\phi$, noting that the higher values at low $\dot{\gamma}$ demonstrate particle migration has taken place;
(f) the pressure $P$ expressed in units of $\eta \kappa/a$;
(g) the shear stress $\sigma_{xy}$ computed from the particle interactions (red points) and by integrating over the left hand side of Eq.~\ref{momBalance} (green points), in the same units as $P$.}
\label{fig1}
\end{figure*}

\paragraph{Simulations details.}
We simulate a mixture of frictionless, non-Brownian spheres of radius $a$ and $1.4a$ mixed in equal number in a periodic box of dimensions $L_x$, $L_y$, $L_z$,
using LAMMPS~\cite{plimptonLAMMPS,ness2023simulating} (see Fig.~\ref{fig1}(a)).
Particles are suspended in a density ($\rho$) matched viscous liquid,
and we impose pairwise contact and hydrodynamic forces as described by Ref.\cite{ChealNessJoR2018}.
Briefly,
the hydrodynamic lubrication force for particles of radius $a_i$ and $a_j$,
with center-to-center vector $\bm{r}_{i,j}$,
is given by $\bm{F}^{h}_{i,j} \sim (1/h)\mathbf{u}_{i,j}$,
where $\mathbf{u}_{i,j}$ is the relative velocity of the particles and $h=(a_i+a_j)-|\bm{r}_{i,j}|$.
$F^{h}_{i,j}$ is not computed for $h > 0.05a$,
and it saturates to $\sim(1/h^c)\mathbf{u}_{i,j}$ for $h \leq h^c$ (with $h^c=0.001a$),
allowing particles to come into contact.
Contact forces arise only when $|\bm{r}_{i,j}|<(a_i+a_j)$ and are given by $\bm{F}^c_{{i,j}} = k\left[(a_i + a_j) - |\bm{r}_{i,j}|\right]\mathbf{n}_{ij}$, where $k$ is a spring constant and $\mathbf{n}_{i,j}=\bm{r}_{i,j}/|\bm{r}_{i,j}|$.
Particles additionally experience dissipative drag due to motion relative to the fluid,
given by $\bm{F}^{d}_i = 6\pi\eta a \left(\bm{u}_i - \bm{u}^\infty(y_i)\right)$,
with $\bm{u}_i$ the velocity of particle $i$ and $\bm{u}^\infty(y_i)$ the liquid streaming velocity at the position of particle $i$.

Flow is generated by specifying $\bm{u}^\infty$ to induce particle motion through drag.
We obtain homogeneous rheology data
for fixed-volume systems of $\phi=0.48$ to $0.65$
by generating simple shear \emph{via}
$\bm{u}^\infty(y)=\dot{\gamma}y\bm{\delta}_x$,
with $y$ the direction of the velocity gradient and $\bm{\delta}_x$ the unit vector along $x$.
We chose our parameters such that $\rho\dot{\gamma}a^2/\eta\ll1$ and $\dot{\gamma}\sqrt{\rho a^3/k}\ll1$, recovering rate-independence~\cite{BoyerGuazelliPouliquenPRL2011}. 
To obtain inhomogeneous flow we specify a spatially dependent liquid velocity as $\bm{u}^\infty(y)=\kappa\sin\left(2\pi y/L_y\right)\bm{\delta}_x$
(see Fig.~\ref{fig1}(b), and the gradient $\dot{\gamma}^\infty$ in Fig.~\ref{fig1}(c)),
and later test the model with $\bm{u}^\infty(y)=\kappa\sin^3(2\pi y/L_y)\bm{\delta}_x$.
We run simulations with $L_y=50a$, $100a$ and $200a$ (with $L_x,L_z=20a$) and systems containing $\mathcal{O}(10^4)$  particles
(we verified that larger systems produce equivalent rheology results).
We simulated systems with mean volume fraction $\bar{\phi}=0.5$ to $0.63$ (achieved by varying the particle number),
and $\kappa$ is a constant with dimensions of velocity,
chosen so that the measured $\rho\dot{\gamma}a^2/\eta$ remains $<0.01$ throughout and particle inertia is negligible.
The stress (a tensor) is computed on a per-particle basis as $\mathbb{\Sigma}_i=\sum_j(\bm{F}_{i,j}^*\otimes\bm{r}_{i,j})$,
counting both contact and hydrodynamic forces.

We aim to compare the spatially-variant values of $J$, $\mu$, $\phi$ and $\Theta$ obtained \emph{via} inhomogeneous flow with the spatially-invariant ones obtained \emph{via} homogeneous flow (the latter follow closely our previous results~\cite{ChealNessJoR2018}). 
Doing so requires computing the variation in $y$ of the stress and velocity fields under inhomogeneous flow,
which we do by binning particle data in blocks of width $a$ and volume $V_b = L_x a L_z$,
with the per-block value of a quantity being simply
the mean of the per-particle quantities of the particles with centers lying therein.
We compute the velocity fluctuation (necessary for calculating the $\Theta$ field) 
of each particle as $\delta u_i = |u_{i,x} - {u}^\dagger_{i,x}|$ where $u_{i,x}$ is the $x$-component of $\bm{u}_i$
and ${u}^\dagger_{i,x}$ is the average $x$ velocity of all particles with centers lying in a narrow window $\pm \epsilon$ (taking $\epsilon=\mathcal{O}(0.1a)$) of $y$,
and we then bin $\delta u_i$ per block.
As all three components of the velocity fluctuations are statistically equivalent
we have used only the $x$ values to compute $\Theta$.
In what follows we report steady state data only~\cite{footnoteCite},
averaging across 6 realizations and at least 500 configurations per realization.

\begin{figure*}
\includegraphics[trim = 0mm 0mm 0mm 0mm, clip,width=0.95\textwidth,page=2]{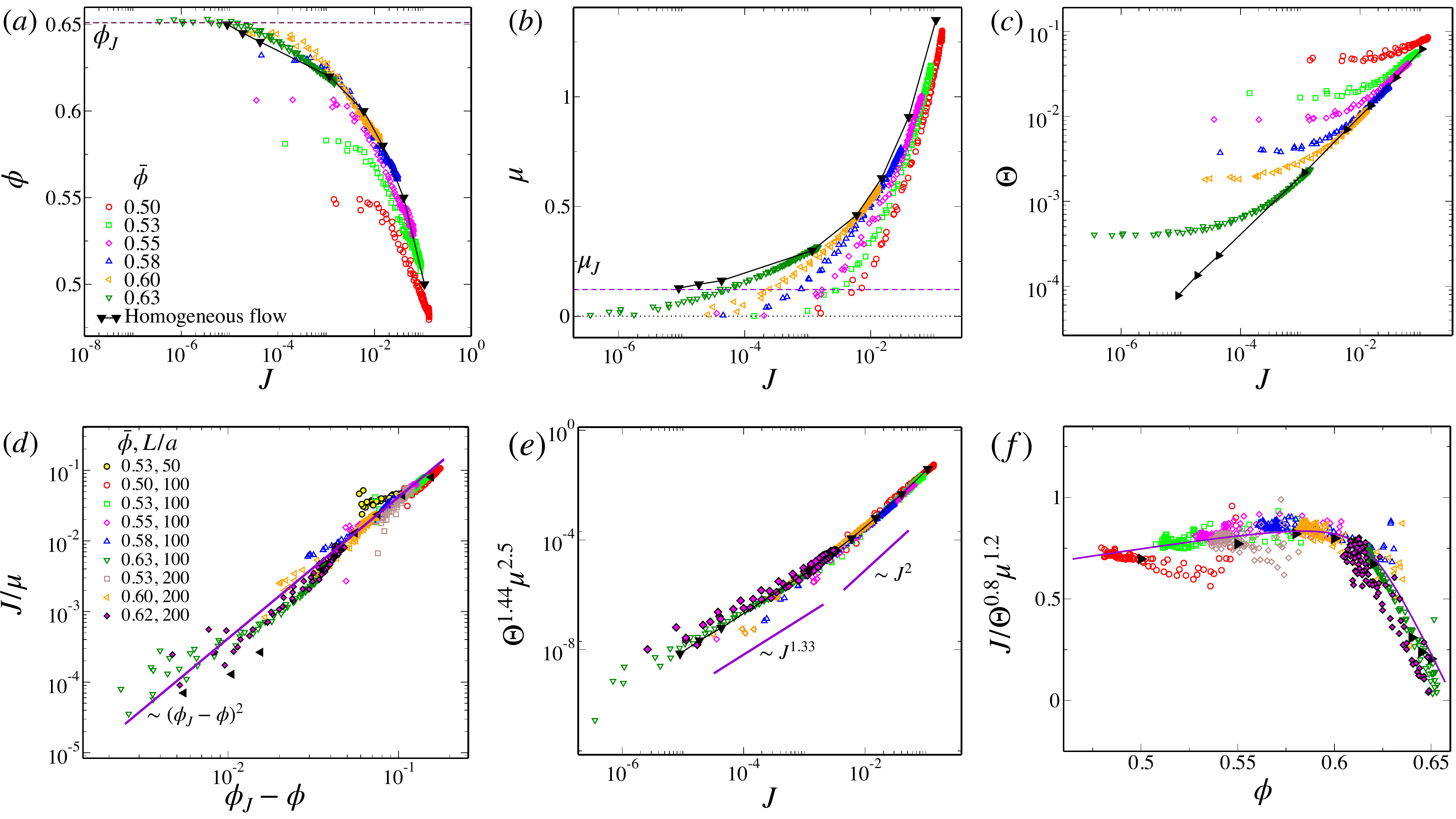}
\caption{
Relations between the dimensionless control parameters.
Shown in the top row are
the relations between the dimensionless viscous number $J$ and
(a) the volume fraction $\phi$ for a range of homogeneous $\phi$ (black data) and inhomogeneous $\bar{\phi}$;
(b) the effective friction coefficient $\mu$ and;
(c) the suspension temperature $\Theta$.
In the bottom row are the collapses using the scaling Eqns.~\ref{scaling1} (d), \ref{scaling2} (e) and \ref{scaling3} (f), for different $\Bar{\phi}$ and $L$.
In (d) we show data for $L/a=50$ to highlight its deviation from the scaling relation.
Black triangles represent homogeneous data (simple shear) and all other points are for inhomogeneous flow at different $\Bar{\phi}$. 
}
\label{fig2}
\end{figure*}

\paragraph{Results.}
Shown in Fig.~\ref{fig1}(b)-(g) are,
respectively,
steady-state profiles in $y$ of the coarse-grained
velocity (in $x$) $u_x$,
shear rate $\dot{\gamma}=\partial u_x/\partial y$,
velocity fluctuations $\delta u$,
volume fraction $\phi$,
pressure $P$ ($=(1/3)\text{Tr}(\mathbb{\Sigma}$)),
and shear stress $\sigma_{xy}$,
for $\Bar{\phi} = 0.60$,
with each plotted point representing a block.
Although at initialisation the particle density is homogeneous (\emph{i.e.} $\phi\neq\phi(y)$),
in the steady state $\phi$
exhibits spatial variation
set up by particle migration to balance the normal stress~\cite{Migration1, Migration2,morris1999curvilinear}.
The velocity profile follows a similar trend to the applied force,
as expected,
but is flattened at the regions of largest $\phi$ 
leading to significant deviations between $\dot{\gamma}$ and $\dot{\gamma}^\infty$.
The pressure becomes spatially uniform, and the shear stress follows the shear rate in sign. 
Since $P$ is
spatially invariant in the steady state,
one can deduce that the variation of the quantities $\eta\dot{\gamma}/P$,
$\sigma_{xy}/P$ and $\eta \delta u/aP$ 
follow
$\dot{\gamma}$,
$\sigma_{xy}$ and $\delta u$ respectively.

We analyse inhomogeneous data by computing the dimensionless control parameters in each block,
defining the scalar shear rate and stress components on the basis of invariants
of the respective tensor quantities so that 
$J,\mu>0$.
 This is done for a range of $\bar{\phi}$,
 with parametric plots of $J(y)$, $\phi(y)$, $\mu(y)$ and $\Theta(y)$ shown in Figs.~\ref{fig2}(a)-(c).
 Each plotted point represents a $y$-coordinate, and colors represent different $\bar{\phi}$.
Shown also (in black) are homogeneous data.
Reading across the data points of a single color from right-to-left represents moves from regions of high-to-low $\dot{\gamma}$ in the inhomogeneous domain.

The homogeneous $\phi(J)$ and $\mu(J)$ relations follow qualitatively the result of~\citet{BoyerGuazelliPouliquenPRL2011},
though our frictionless particles render $\phi_J$ and $\mu_J$ dissimilar.
$\Theta(J)$ follows a power-law relation,
as in dry granular matter~\cite{KimKamrin2020PRL} though with a different exponent.
In general
large-$J$ inhomogeneous data approximately match homogeneous data,
though they deviate with decreasing $J$
demonstrating the shortcomings of the existing constitutive laws.

With the help of scaling theory, we next attempt to find constitutive laws that simultaneously describe the rheology under homogeneous and inhomogeneous flow.
We focus first on how the inverse viscosity
$J/\mu=\eta\dot{\gamma}/\sigma_{xy}$ vanishes as $\phi$ approaches the jamming point $\phi_J$.
This trend is followed by all the homogeneous and inhomogeneous simulations, leading to our first scaling relation
\begin{equation}
J/\mu = \alpha\left(\phi_J - \phi\right)^2\text{,}
\label{scaling1}
\end{equation}
plotted in Fig.~\ref{fig2}(d) with $\alpha=4.1$ and $\phi_J=0.6555$.

The next scaling relation is motivated by~\citet{KimKamrin2020PRL}.
In homogeneous flow,
within the range of our data we find $\mu^{2.5} \sim J$ (Fig.~\ref{fig2}(b)) and $\Theta^{1.44} \sim J$ (Fig.~\ref{fig2}(c)).
Since for the range of $\bar{\phi}$ explored here
inhomogeneous data follow homogeneous laws at large $J$,
we expect a scaling of the form $\mu^{2.5}\Theta^{1.44} \sim F_1(J)$.
Indeed this results in a good collapse as shown in Fig.~\ref{fig2}(e),
in which data are described by the relation
\begin{align}
\Theta^{1.44}\mu^{2.5} = \begin{cases}
\beta J^2 & \text{if $J > 10^{-3}$}; \\
\vartheta J^{1.33} & \text{if $J \le 10^{-3}$};\\
\end{cases}
\label{scaling2}
\end{align}
with $\beta=3$ and $\vartheta=0.06$.

The final scaling relation is motivated by the relation between granular fluidity and $\phi$ reported for dry granular matter.
\citet{ZhangKamrinPRL2017} write a non-dimensional granular fluidity $\tilde{g} = gd/\delta u$, where $g = \dot{\gamma}/\mu$, and $d$ is the spatial dimension.
We define an equivalent quantity in terms of the previously discussed dimensionless numbers, namely $J/\mu\Theta$,
though we find a better collapse is achieved through a change to the exponents as
\begin{equation}
\frac{J}{\Theta^{0.8}\mu^{1.2}} = F_{2}(\phi)\text{,}
\label{scaling3}
\end{equation}  
with $F_2(\phi) = \epsilon\left[(\phi - \phi_m) + \sqrt{(\phi - \phi_m)^2 + \zeta}\right] + \lambda\phi$
(see Fig.~\ref{fig2}(f))
and
$\epsilon=-10.98$,
$\phi_m=0.618$,
$\zeta=0.0004$
and $\lambda=1.533$.
We thus have three scaling relations, Eqs.~\ref{scaling1},~\ref{scaling2} and~\ref{scaling3},
that relate $\phi$, $J$, $\mu$ and $\Theta$.
The collapse appears poorer for $\bar{\phi}=0.5$ (Fig.~\ref{fig2}(f))
and $L/a=50$ (Fig.~\ref{fig2}(d)), indicating limits to the range of applicability.
An issue in the former case may be that our simplified hydrodynamics,
accounting only for lubrication,
becomes nonphysical at lower $\phi$ and that a more highly resolved fluid field is required.

Given a profile of one of the dimensionless numbers,
one could therefore fully characterise the rheology of the system.
In our simulations,
however,
the only known input is the externally applied force,
which we recall is defined through $\bm{u}^\infty$.
To use the scaling relations we need to establish another relation that can provide us one of these dimensionless numbers from the knowledge of the applied force profile.
Considering the inertia-free momentum balance $\nabla\cdot\mathbb{\Sigma} = -\bm{f}$ per unit volume,
we can write the following equation for the $k^{th}$ block of the simulation cell (which we verified in Fig.~\ref{fig1}(g)):
\begin{equation}
N_k6\pi\eta a\left[Au_{x,k}^\infty -  u_{x,k} \right] = -\left( \frac{\partial\sigma_{xy,k}}{\partial y} \right) V_b\text{.}
\label{momBalance}
\end{equation} 

\begin{figure}
\includegraphics[trim = 0mm 0mm 271mm 0mm, clip,width=0.48\textwidth,page=3]{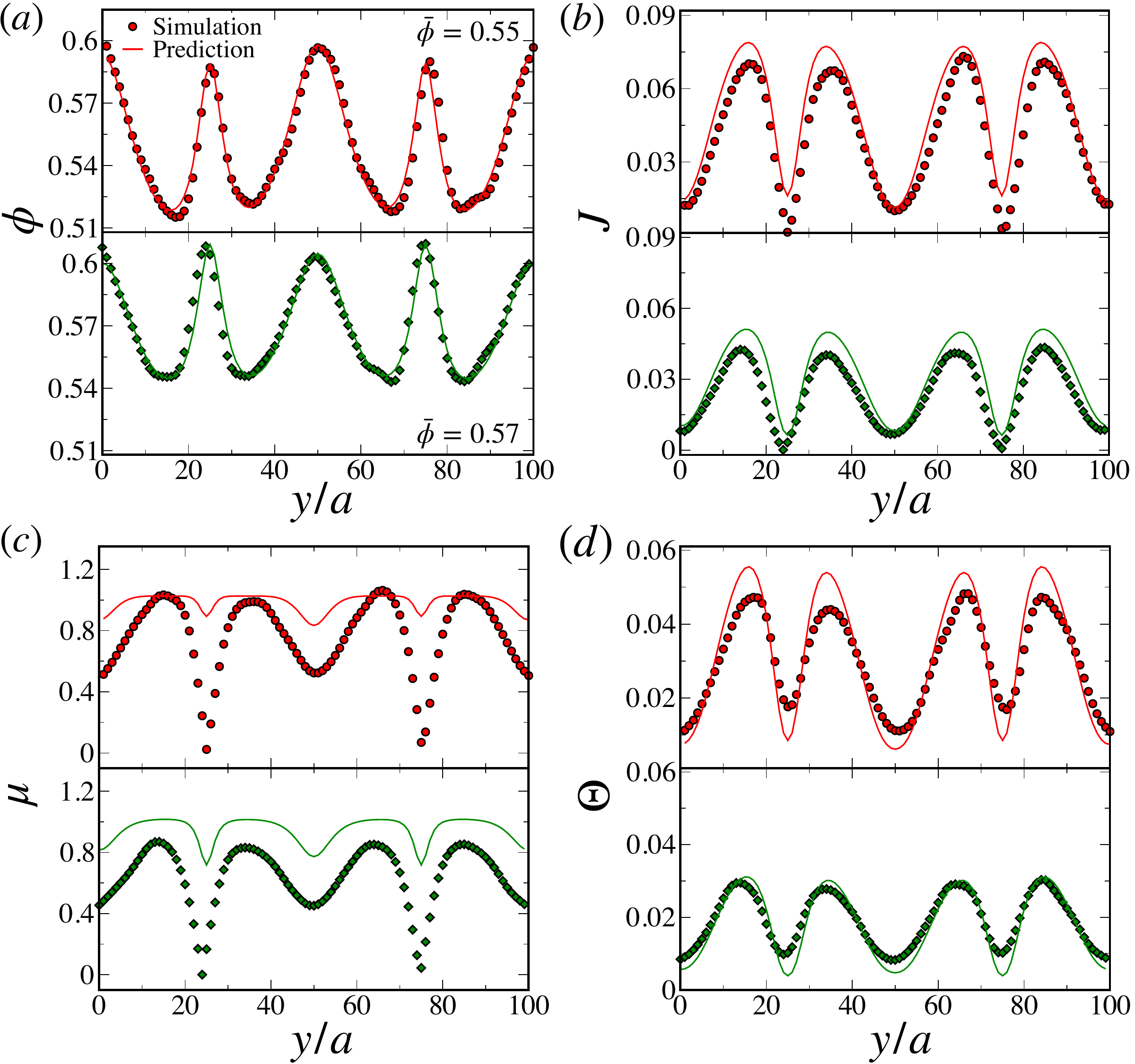}
\caption{Predictions of the scaling relations against simulation data not used for obtaining the scaling exponents, with $\bm{u}^\infty(y)=\kappa\sin^3(2\pi y/L_y)\bm{\delta}_x$.
Shown are (a) the volume fraction $\phi$;
(b) the viscous number $J$;
(c) the effective friction coefficient $\mu$;
and (d) the suspension temperature $\Theta$,
with predictions given by solid lines and simulation data in points, for $\Bar{\phi} = 0.55$ (red) and $0.57$ (green).}
\label{fig3}
\end{figure}

\noindent Here $N_k$, $u_{x,k}^\infty$, $ u_{x,k}$
and $\sigma_{xy,k}$ are 
the particle number in the block,
the liquid streaming velocity at the centre of the block,
and the particle velocity and stress averaged over the block, which has volume $V_b$.
$A$ is an order unity quantity
necessary to account for small variations in $u_{x,k}^\infty$ across the block.
The first term of Eq.~\ref{momBalance} represents the net applied force and the second represents the net viscous force exerted by the fluid due to drag. 
The resultant of these is balanced by the net stress gradient inside the block.
Using the definition of our dimensionless numbers,  Eq.~\ref{momBalance} can be rewritten for the streaming velocity at $y$ as

\begin{equation}
u_{x}^{\prime\infty}(y) = \left[\int_0^y\frac{1}{a}J^*(y')dy' - \frac{2a}{9\phi(y)}\left( \frac{\partial\mu^*(y)}{\partial y} \right)\right]\text{,}
\label{momBalanceDimLess}
\end{equation}
with $u_{x}^{\prime\infty}(y)=u_{x}^{\infty}(y)\eta A/aP$
and asterisks representing multiplication by $\mathrm{sgn}(\dot{\gamma}^\infty(y))$,
noting that $P$ is uniform at steady state
and using $\phi(y) = (4/3)\pi a^3 N(y)/V_b$, acknowledging our earlier comment about phase separation~\cite{footnoteCite}.
Equation~\ref{momBalanceDimLess} thus relates the externally applied liquid flow field to the profiles of $J$, $\mu$ and $\phi$.

For a known $\bm{u}^\infty$
we solve Eqs.~\ref{scaling1}, \ref{scaling2}, \ref{scaling3} and~\ref{momBalanceDimLess}
numerically in the following way.
We first guess a $\phi\left(y\right)$ profile
by assuming accumulation at points where the spatial derivative of the imposed force vanishes,
starting with a simple form as
$\phi(y) = \sum_{j=1}^{n_p} a_{j}/[(y - y^0_j)^2 + b^2_{j}] + \phi_0$\text{,}
with mass conserved through
$\bar{\phi} = \frac{1}{L_y}\int_0^{L_y} \phi\left(y\right)dy$.
Here $y^0_j$ are the coordinates of the point where the first derivative of the applied force vanishes,
$n_p$ is the number of such points and $b_j$ is the width of the Lorentzian function peaked at $y^0_j$.
We then compute directly $J$, $\mu$ and $\Theta$ using Eqs.~\ref{scaling1}, \ref{scaling2} and \ref{scaling3},
before attempting to balance Eq.~\ref{momBalanceDimLess}.
The imbalance of Eq.~\ref{momBalanceDimLess} reflects the accuracy of our guess.
We refine $\phi(y)$ by tuning $\phi_0$, $a_j$ and $b_j$ until Eq.~\ref{momBalanceDimLess} is satisfied (up to some tolerance).         
Shown in Fig.~\ref{fig3}
are predicted results compared against `unseen' simulation data (\emph{i.e.} data not used to obtain the scaling exponents) with $\bar{\phi}=0.55$, $0.57$ and $\bm{u}^\infty(y)=\kappa\sin^3(2\pi y/L_y)\bm{\delta}_x$,
demonstrating the degree of success of the scaling relations for predicting $y$-profiles of $\phi$, $J$, $\mu$ and $\Theta$.
Considering the highly non-linear nature of the scaling relations, the quality of the predictions is reasonably good.
 
\paragraph{Conclusions.} Using particle-based simulation we seek universality
amongst flows of dense, frictionless suspensions.
Along with canonical suspension rheology control parameters $\phi$, $J$ and $\mu$,
we introduce a fourth quantity $\Theta$ characterising velocity fluctuations,
inspired by recent studies in dry granular physics~\cite{KimKamrin2020PRL}.
We find a trio of scaling relations among these quantities that collapse data for homogeneous and inhomogeneous flow.
Utilising a momentum balance we show that from knowledge of the externally applied force,
one can use the relations
to predict the features of a general inhomogeneous flow.
Our work raises manifold avenues for future work.
In particular,
the microscopic origin of the exponents is not understood,
nor is their generalisation to
the broader class of suspensions that includes polydisperse particles (for which colloidal forces may become relevant~\cite{li2023simulating}), non-spheres and other complexities.
Meanwhile the question of a diverging lengthscale
---apparently a staple of non-local rheology in dry granular matter~\cite{TrulssonNolocal, KamrinKovalPRL2012,tang2018nonlocal}---
remains open.
Computing a granular fluidity field from our data,
we find,
similar to~\cite{TigheNonLocalPRL},
no divergence in the characteristic lengthscale,
which remains $\mathcal{O}(a)$ everywhere.
This raises an important open question regarding what are the minimal conditions required for a diverging lengthscale in inhomogeneous particulate flows. 

B.P.B. acknowledges support from the Leverhulme Trust under Research Project Grant RPG-2022-095;
C.N. acknowledges support from the Royal Academy of Engineering under the Research Fellowship scheme.
We thank Ken Kamrin, Martin Trulsson, Mehdi Bouzid, Romain Mari and Jeff Morris for useful discussions.

\bibliography{ALL}

\begin{thebibliography}{35}
\bibitem [{foo()}]{footnoteCite}%
  \BibitemOpen
  \href@noop {} {\bibinfo  {journal} {There is in fact a very slow phase
  separation of small and large particles driven by the shear rate gradient, so
  we keep run times sufficiently short that no significant re- distribution
  occurs and our data represent uniform mix- tures}\ }\BibitemShut {NoStop}%
\end{thebibliography}

\end{document}